\documentclass[aps,twocolumn,groupedaddress]{revtex4-1}
\usepackage{graphicx}
\usepackage{color}
\usepackage{amssymb}
\begin{document}
\title{An elastic-network based local molecular field analysis of zinc-finger proteins}
\author{Purushottam D. Dixit}
\author{D. Asthagiri} \thanks{Corresponding author email: dilipa@jhu.edu}
\affiliation{Chemical and Biomolecular Engineering, Johns Hopkins University, Baltimore, MD 21218} 
\begin{abstract}
We study two designed and one natural zinc-finger peptide each with the Cys$_2$His$_2$ (CCHH) type of metal binding motif. 
In the approach we have developed, we describe the role of the protein and solvent outside the Zn(II)-CCHH metal-residue 
cluster by a molecular field represented by generalized harmonic restraints. The strength of the field is adjusted to reproduce the binding energy distribution of the metal with the cluster obtained in a reference all-atom simulation with empirical potentials. The quadratic field allows us to investigate analytically the protein restraints on the binding site in terms of its eigenmodes. Examining these eigenmodes suggests, consistent with experimental observations, the importance of the first histidine (H) in the CCHH
cluster in metal binding. Further, the eigenvalues corresponding to these modes also indicate that the designed proteins form 
a tighter complex with the metal.  We find that the bulk protein and solvent response tends to destabilize metal-binding, emphasizing that the favorable energetics of metal-residue interaction is necessary to drive folding in this system.  
The representation of the bulk protein and solvent response by a local field allows us to perform 
Monte Carlo simulations of the metal-residue cluster using quantum-chemical approaches, here using a semi-empirical Hamiltonian. For configurations sampled
from this simulation, we study the free energy of replacing Zn$^{2+}$ with Fe$^{2+}$, Co$^{2+}$, and Ni$^{2+}$ using density functional theory. The calculated selectivities are in fair agreement with experimental results. 
\end{abstract}
\maketitle
\section{Introduction\label{sec-intro}}
An important post-translational modification of proteins involves incorporating metal ions into the protein structure~\cite{gray:ic04}. In many of these instances, metal-binding stabilizes the folded structure or helps fold a previously unstructured or partially structured peptide.  The versatility of metals as stabilizers or modifiers of protein structure is principally due to the nature of their interactions with {amino acid residues}: substantially strong on a thermal energy scale and chemically intricate~\cite{bergbook}. 

The prevalence of metalloproteins and the growing appreciation of metal-induced (mis)folding in diseases, 
for example, see Refs.\ \cite{Millhauser:acc04,Wilson:qrb04,Adlard:jad06}, makes obtaining a molecular level understanding of the role of metals in protein structure and function of unquestionable importance. But the intricacies of metal-protein interactions makes this a formidable challenge to current theory and simulation approaches. In a step towards this larger challenge, here we address the role of material outside the first-shell of the metal in metal-binding and selectivity in a zinc-finger protein. 

A satisfactory description of metal-protein interactions requires quantum chemical calculations, and these calculations, especially at a high-level of theory, are computationally demanding. Hence quantum chemical calculations are limited to a small group of residues surrounding the metal ion.  The effect of the remaining protein and solvent medium on the structure and dynamics of the metal-residue cluster is typically described in one of three ways (for example, see Refs.\ \cite{siegbahn:rev99,ryde:cocb03,truhlar:tca07,thiel:angew09}): the medium is entirely ignored, effectively simulating the cluster in {\em vacuo\/}; the medium is described as a continuum with a dielectric constant; and in the most sophisticated of methods, the medium is described using empirical forcefields. Among these alternatives, only the last method explicitly accounts for the role of the bulk in modulating the architecture and dynamics of the cluster. 

{The approach we present is a rigorous reduction in the degrees of freedom of the system and has the aim of 
understanding the role of the medium outside the defined metal-residue cluster: the cluster is described in atomic detail and 
its structure and dynamics are influenced by the medium whose effect is described by a local molecular field.} 
The present approach is inspired by a recent development in the theory of liquids~\cite{lrp:hspre}. The essential idea in that development is to describe the role of the medium external to a defined inner-shell~\cite{lrp:hspre} around a solute  by a molecular field whose strength is adjusted to satisfy suitable consistency conditions, such as the mean density of the inner-shell. 

Here we obtain the local molecular field by describing the bulk protein outside the cluster 
as an elastic medium ~\cite{brooks:bj06,brooks:ps07,brooks:jcp08,bahar:bj01,bahar:cosb05,micheletti:ps04,micheletti:prl07,micheletti:jpr07,micheletti:ps09}.  
 This development allows us to separate the system Hamiltonian into that for the 
metal-residue cluster and the remainder. We then integrate out the bulk degrees of freedom
 to obtain a molecular field acting on the cluster. The strength of the field is 
adjusted self-consistently such that the binding energy distribution of the metal with the local
residues reproduces the binding energy distribution obtained in a simulation treating the bulk 
atomically as well. (We focus on the binding energy distribution, as this is the most relevant quantity for understanding the thermodynamics of metal binding to the protein~\cite{asthagiri:jcp06,lrp:cpms,dixit:bj09,dixit:bj11}.) In this initial study, the reference all-atom simulation and the simulations to determine the strength of the molecular field are all performed using empirical forcefields. With the local molecular field determined at the coarse level, we study the metal-residue cluster using higher level 
methods, here using semi-empirical and density functional methods with large basis sets. 

We use the local molecular field approach to study Zn$^{2+}$ binding to a zinc finger domain. Zinc finger domains are widely distributed in cellular systems, most notably in the transcription factor assembly~\cite{Klug:1995p3038,berg:sc96,Wolfe:2000p3044}. The isolated zinc-finger domain is partially unstructured in the absence of the metal and achieves the correct folded conformation only upon binding the metal \cite{Frankel:1987p3085,Michael:1992p3116,berg:arbbs97}.  In the folded state, Zn$^{2+}$ is coordinated tetrahedrally to n-cysteine and 4-n histidine residues where n can be 2, 3 or 4. In the system we study, n is 2.

The rest of the article is organized as follows. In Sec.~\ref{sec-theory} we derive the local molecular field. Section~\ref{sec-methods} collects details of the molecular simulations. Discussions and conclusions follow results presented in Sec.~\ref{sec-results}.

\section{Theory}\label{sec-theory}
Consider the metal-bound protein in a solvent medium.  We denote the conformational degrees of the protein 
by $\mathcal X = (\mathcal{X}_1,\mathcal{X}_2)$, where $\mathcal X_1$ is the metal plus
the neighboring {amino acid residues} and $\mathcal X_2$ is the remainder of the protein. The solvent degrees
are given by $\mathcal X_s$. 

In the canonical ensemble, for a given protein coordinate $\mathcal{X}$, we can formally integrate over the solvent degrees of freedom and write the effective potential on the protein --- the potential of mean force --- as 
\begin{eqnarray}
\mathcal U(\mathcal X_1, \mathcal X_2; \beta)  & = & U_{1}(\mathcal X_1) +   U_{12}(\mathcal X_1 , \mathcal X_2)  +  U_{2}(\mathcal X_2) +  \eta(\mathcal{X}_1, \mathcal{X}_2; \beta) \label{eq:expand}
\end{eqnarray}
$U_1(\mathcal X_1)$, $U_{12}(\mathcal X_1, \mathcal X_2)$, and $U_2(\mathcal X_2)$  are site-site, site-bulk, and 
bulk-bulk interactions and $\eta(\mathcal{X}_1,\mathcal{X}_2; \beta)$ is the solvent response.  (Note that, in principle, such a decomposition can always be made.) $\beta = 1/k_{\rm B}T$, where $T$ is the temperature and $k_{\rm B}$ is the Boltzmann constant. We indicate the temperature dependence of $\eta$, a thermally averaged quantity, to emphasize the distinction from the potentials $U_1$, $U_2$, and $U_{12}$. 

In QM/MM approaches~\cite{ryde:cocb03,truhlar:tca07,thiel:angew09,karplus:jpca00}, $U_1$ is described quantum mechanically and $U_2$ and $U_{12}$ are described using molecular mechanics. Here we approximate the latter two quantities {together with the response of the solvent} by generalized harmonic restraints around the equilibrium structure of the protein. Our {\em ansatz} is 
\begin{eqnarray}
\mathcal   U(\mathcal X_1, \mathcal X_2; \beta) \approx U_1(\mathcal X_1) + \left [ \delta\mathcal  X_1, \delta \mathcal X_2\right ] \alpha\mathcal H(\beta) \left [ \begin{array}{c} \delta \mathcal X_1\\ \delta \mathcal X_2 \end{array}\right ] 
\label{eq:approx}
\end{eqnarray}
where
\begin{eqnarray}
	\delta \mathcal X_i &=& \mathcal X_i - \mathcal X_{i,p} \;\; {\rm for\;} i = 1, 2. 
\end{eqnarray}
Here, $\mathcal X_{i,p}$ for $i=1,2$ are the reference coordinates for the binding site and the protein medium obtained from the three dimensional structure of the protein.  The approximation made in Eq.~\ref{eq:approx} is physically motivated and attempts to describe the viscous damping of protein oscillations as harmonic fluctuations~\cite{micheletti:ps04}, but explicit 
solvent-binding site interactions are neglected. We anticipate that these long-range contributions can be described using mean-field models (such as dielectric models) and that they will not contribute to discriminating between {Zn$^{2+}$} and a competing metal bound at the site.

Our plan is to obtain the quadratic Hamiltonian, $\mathcal{H}$, using contact topology based potentials. Since these models are valid only up to a proportionality constant~\cite{brooks:bj06,brooks:ps07,brooks:jcp08,bahar:bj01,bahar:cosb05,micheletti:ps04,micheletti:ps09},  to construct a physical potential energy function, we introduce a coupling constant, $\alpha$, that fixes the strength of the harmonic interaction. 

The matrix $\mathcal H$, suppressing the dependence on $\beta$ for simplicity, is expanded as
\begin{eqnarray}
 \mathcal H =  \left [ \begin{array}{cc} \mathcal H_1& G \\ G^{T} & \mathcal H_2 \end{array}\right ] .
\label{eq:hmat}
 \end{eqnarray}
Here $\mathcal H_1$ is a diagonal matrix: no two site-atoms couple through $\mathcal H_1$  as those interactions are explicitly described in $\mathcal U_1$.  (In this regard, the present development differs from those presented earlier~\cite{brooks:bj06,brooks:jcp08,ming:prl05}.) The matrix $\mathcal G$ couples the site atoms to the bulk, and the matrix $\mathcal H_2$ couples the bulk atoms with each other. 

Under the approximations noted above, for the system modeled by $U(\mathcal X_1, \mathcal X_2; \beta)$, the excess Helmholtz free energy, $\mathcal A^{\rm ex}$, is given by
\begin{eqnarray}
e^{-\beta \mathcal A^{\rm ex}} &=& \int e^{-\beta \mathcal U(\mathcal X_1,\mathcal X_2;\beta)} d\mathcal X_1d\mathcal X_2.
\label{eq:canon}
\end{eqnarray}
Since the bulk protein coordinates $\mathcal X_2$ appear quadratically, following~\cite{ming:prl05}, we 
integrate over $\mathcal X_2$ and get
\begin{eqnarray}
e^{-\beta \mathcal A^{\rm ex}} &=&\int e^{-\beta \left( U_1(\mathcal X_1) +  \delta \mathcal X_1^{T} \alpha \mathcal H_{\rm site} \delta \mathcal X_1\right)}\nonumber \\
	&\equiv& \int e^{-\beta \mathcal U(\mathcal X_1;\beta)}d\mathcal X_1
\label{eq:aex}
\end{eqnarray} 
where
\begin{eqnarray}
\mathcal H_{\rm site} &=& \mathcal H_1 - G \mathcal H_2^{-1} G^{T}\label{eq:wall} .
\end{eqnarray}
 Thus, under the approximations noted above, the effective potential for the site is given by 
\begin{eqnarray}
\mathcal U(\mathcal X_1; \beta) &=& U_1(\mathcal X_1) +  \delta \mathcal X_1^{T} \alpha \mathcal H_{\rm site} \delta \mathcal X_1 \nonumber \\
 &=& U_1(\mathcal X_1) +  \phi_m(\mathcal X_1;\alpha) ,
\label{eq:pot}
\end{eqnarray}
where in addition to the site-site interaction potential, $U_1(\mathcal X_1)$, the effective potential contains a  local molecular field, $\phi_m(\mathcal X_1;\alpha) = \delta \mathcal X_1^{T} \alpha \mathcal H_{\rm site} \delta\mathcal X_1$,  that describes the effect of the bulk protein and solvent damping on the site atoms.  
(For notational simplicity, the temperature dependence on $\phi_m$ is not explicitly shown.)  
The field $\phi_m$ acts as a restraint that limits the deflections of the binding site away from some reference state. (A suitable reference state can be the PDB structure.)  For $\alpha = 0$, there is no coupling between the site and the bulk, and the model reduces to a metal-residue cluster in vacuum. 

\subsection{Selectivity of the binding site}
The selectivity of the zinc finger peptide for Zn$^{2+}$ over another transition metal X$^{2+}$ (X$^{2+}$ = Co$^{2+}$, Fe$^{2+}$, and Ni$^{2+}$) is determined by 
\begin{eqnarray}
\Delta \mu^{\rm ex} &=& [\mu^{\rm ex}_{\rm X^{2+}} - \mu^{\rm ex}_{\rm Zn^{2+}}]({\rm S}) - [\mu^{\rm ex}_{\rm X^{2+}} - \mu^{\rm ex}_{\rm Zn^{2+}}]({\rm aq})  \nonumber \\
      &=& \Delta \mu^{\rm ex}({\rm S}) - \Delta \mu^{\rm ex}({\rm aq}) \, ,\label{eq:delmu} 
\end{eqnarray}
where $\mu^{\rm ex}$(aq) is the change in hydration free energies  and 
$\mu ^{\rm ex}$(S) is the corresponding quantity in the protein.  It is understood that a common reference state is used in defining $\mu^{\rm ex}_{\rm X^{2+}}$ in water and in the protein \cite{asthagiri:jcp06,dixit:bj09}. 

Calculating $\Delta \mu^{\rm ex}$(S) presents significant challenges, including  the need to describe interactions
quantum mechanically together with sampling binding site conformations. It is here that the reduction in degrees of freedom made possible by the effective potential (Eq.~\ref{eq:pot}) proves helpful.

Following Eq.~\ref{eq:aex}, $\Delta\mu^{\rm ex}(\rm S)$ is given by \cite{asthagiri:jcp06,lrp:cpms,lrp:book}
\begin{eqnarray}
	e^{-\beta \Delta \mu^{\rm ex}(\rm S)} 	&=& \langle e^{-\beta \Delta U_1 (\mathcal X_1)}\cdot e^{-\beta \Delta \phi_m}\rangle_{\rm Zn^{2+}} \nonumber \\ 
	&\approx&  \langle e^{-\beta \Delta U_1 (\mathcal X_1)}\rangle_{\rm Zn^{2+}}\, , \label{eq:apmu}
\end{eqnarray}
where $\mathcal U_{\rm X^{2+}}(\mathcal X_1;\beta)$ is the effective potential (Eq.~\ref{eq:pot}) with
$\rm X^{2+}$ in the site, $\Delta U_1 = U_{1,\rm X^{2+}} - U_{1,\rm Zn^{2+}}$, $\Delta \phi_m = \phi_{m,{\rm X^{2+}}} - \phi_{m {\rm Zn^{2+}}}$, and $\langle \ldots\rangle_{\rm Zn^{2+}}$ indicates canonical averaging with {Zn$^{2+}$} bound to the protein.
In Eq.~\ref{eq:apmu}, by ignoring $\Delta\phi_m$ we are assuming that the response of the material outside the defined metal-residue cluster is the same for both X$^{2+}$ and Zn$^{2+}$ and thus the contribution to the selectivity free energy arises solely from interactions within the binding site. (We comment on this approximation below.) 

\subsection{Coupling constant $\alpha$}
The excess chemical potential, $\mu^{\rm ex}_{\rm Zn^{2+}}$, is 
given by the potential distribution theorem~\cite{widom:jpc82,lrp:cpms,lrp:book}
\begin{eqnarray}
\mu^{\rm ex}_{\rm Zn^{2+}} = k_{\rm B} T \log \left [ \int_{-\infty}^{\infty} e^{\beta \varepsilon} P(\varepsilon) d\varepsilon \right ] ,
\label{eq:pdt}
\end{eqnarray}
where $P(\varepsilon)$ is the probability distribution of the binding energy of the metal ion with the surrounding material.
Since $\mu^{\rm ex}_{\rm Zn^{2+}}$ is the essential quantity characterizing the 
thermodynamics of Zn$^{2+}$ binding to the site,  we choose $\alpha$ such that the energy distribution from the approximate model (Eq.~\ref{eq:pot}) reproduces the binding energy distribution of Zn$^{2+}$ with the site from all-atom molecular simulations. It is most economical computationally, if 
these simulations are based either  on an empirical potential energy function or a QM/MM approach with 
a less expensive quantum mechanical model. (Here we use an empirical potential energy function.) 
Then with the molecular field based on a coarse energy function, one can refine the description of the site with high-level quantum mechanical calculations in the presence of the molecular field.

To obtain $\alpha$, we minimize the Kullback-Liebler divergence~\cite{kl:ams51} between the distribution, $ P_s(\varepsilon)$, obtained with simulation using a coarse model, and the distribution, $P(\varepsilon;\alpha)$, obtained from simulations of the site with the effective potential (Eq.~\ref{eq:pot}). Due to the exponential weighting $e^{\beta \varepsilon}$, the high-$\varepsilon$ tail of the energy distribution $P(\varepsilon)$ more sensitively determines the excess free energy in Eq.~\ref{eq:pdt}. Hence we require simulations with the model Hamiltonian (Eq.~\ref{eq:pot}) to reproduce the high-$\varepsilon$ tail. Thus, we restrict our attention to energy values $\varepsilon \geq \bar \varepsilon$, where the mean binding energy is $\bar \varepsilon$. $\alpha$ is then obtained by minimizing 
\begin{eqnarray}
{\Delta (\alpha)} = {\log \int_{\bar \varepsilon}^{\infty}P_s(\varepsilon) \log \frac{ P_s(\varepsilon)}{P(\varepsilon;\alpha)} d\varepsilon .} \label{eq:kl} 
\end{eqnarray}

\section{Methods\label{sec-methods}}
\subsection{Elastic network model}
We construct the quadratic Hamiltonan  ($\mathcal{H}$) based on the Gaussian network model described in Ref.~\cite{micheletti:ps04}. The interaction energy is expanded around the equilibrium structure of the protein in a Taylor series and only terms to second order are retained. 

Denoting the equilibrium distance between particle $i$ and $j$ as $r_{ij}$, the deviated distance by $d_{ij}$, and the deviation by $x_{ij} (= d_{ij} - r_{ij})$, we have  
\begin{eqnarray}
 U(d_{ij}) \approx U(r_{ij}) + \frac{U''(r_{ij})}{2}\sum_{\mu , \nu} \frac{r_{ij}^{\mu}r_{ij}^{\nu}}{r_{ij}^{2}}x_{ij}^{\mu}x_{ij}^{\nu} . 
\end{eqnarray}
$\mu$,$\nu$ are the labels for Cartesian components $x$, $y$, and $z$. $U''(r_{ij})$ is the second derivative of the potential energy function. For simplicity, we assume that the second derivative is equal for all the interactions and that only particles within 6~{\AA} of each other interact. This cutoff value follows earlier studies using elastic network models and other potential functions dependent on topological contact maps~\cite{brooks:bj06,brooks:ps07,brooks:jcp08,bahar:bj01,bahar:cosb05,micheletti:ps04,micheletti:ps09}. We do not include $U_1$ type interactions in constructing $\mathcal H$ as those interactions 
are dealt  in atomic detail. Further, the inversion of $\mathcal H_2$ (Eq.~\ref{eq:wall}) is defined only within the space of eigenvectors with non-zero eigenvalues; the remaining six  zero-modes correspond to rotations and translations and do not contribute to the potential energy. 

\subsection{Molecular dynamics simulation}
We consider three different zinc finger proteins to evaluate our model: the consensus peptide (CP1)~(\cite{berg:arbbs97}; PDB ID: 1MEY), the peptide (YTA) based on the human zinc finger protein 32 (PDB ID: 2YTA), and TF3, and the transcription factor IIIA (\cite{wright:nsb97}; PDB ID: 1TF3). Of these CP1 and YTA are designed peptides. Residues between 139 and 162 forming the zinc-finger domain in YTA are used in the simulations; residues outside the zinc-finger domain are not considered.  The zinc-atom is coordinated by two cysteine thiolates and two histidines. We use the CHARMM27 forcefield~\cite{charmm:jpcb98} for all the {amino acid residues}, with the thiolate partial charges from Refs.~\cite{nilsson:thiolate01,simonson:jmgm06}. Simulations are performed with NAMD2~\cite{namd:cc05}.

We use the dummy cation Zn$^{2+}$ model developed in Ref.~\cite{pang:psfg01}. The key feature of this model is the presence of four dummy sites disposed tetrahedrally at a distance of 0.9~{\AA} from a central atom. The dummy sites have a partial charge of $0.5e$ but no size. The presence of dummy sites proves helpful in maintaining the four-coordinate state of the metal in extended simulations  \cite{pang:psfg01}.

Each of the proteins CP1, YTA, and TF3 are solvated in TIP3P~\cite{tip32,tip3mod} water molecules using the solvate module in VMD~\cite{vmd:jmg96}. After an initial minimization, the systems are equilibrated for 1~ns at 298.15 K and 1 bar followed by a 3~ns production phase. Configurations are saved every 0.1~ps for analysis.  The temperature is maintained by a Langevin thermostat  while pressure is maintained using a Langevin barostat~\cite{feller:jcp95}. The reference binding energy distribution $P_s(\varepsilon)$ (Eq.~\ref{eq:kl}) was calculated using code developed in-house. 
\subsection{Monte Carlo simulations}
\begin{center}
\begin{figure}
\includegraphics[scale=1]{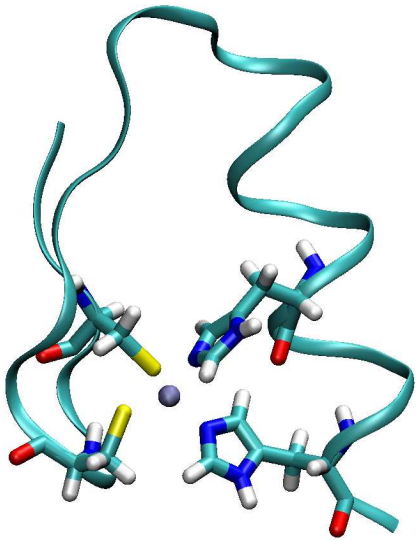}
\caption{The zinc finger domain (pdb ID:2YTA) comprises the $\alpha$ helix and the $\beta$ sheet. The binding site residues (Cys and His) are shown as stick figures. These residues and the metal (sphere) totaling
59 particles are described by the site coordinates $\mathcal X_1$ (Eqs.~\ref{eq:expand} and~\ref{eq:pot}) and are throughout represented in atomic detail.} \label{fg:znf}
\end{figure}
\end{center}
The $\mathcal X_1$ coordinates describe the Zn$^{2+}$ ion and the 4 binding site residues, in all 59 atoms (Figure~\ref{fg:znf}). 
{We calculate $U_1(\mathcal X_1)$ (Eq.~\ref{eq:pot}) using the CHARMM \cite{charmm:jpcb98}  potential energy function
that we have implemented in a collection of Fortran codes.}
{To calculate the field contribution in Eq.~\ref{eq:pot}, we find the deflection of a configuration from the protein reference 
structure,} ${\delta \mathcal X_1} = {\mathcal X_1 - \mathcal X_{1,p}\, ,}$ {and obtain $\phi_{m}(\mathcal {X}_1; \alpha)$ form the known matrix $H_{\rm site}$ using} 
\begin{eqnarray*}
{\phi_m(\mathcal X_1;\beta)} = {\delta \mathcal X_1 \alpha H_{\rm site} \delta \mathcal X_1}\, .
\end{eqnarray*}

For each $\alpha$, the simulations consist of $15\times 10^6$ sweeps of equilibration followed by $15\times 10^6$ sweeps of production. {Every sweep comprises three sets of moves: 1) displacement of each particle;  2) rigid-body displacement
of each amino acid residue; and 3) rigid-body rotation of each amino acid residue. Each of these moves is performed with a probability of $0.5$. 
Displacements are made along randomly chosen directions and rotations are made along each axis using randomly chosen angles. 
The standard Metropolis criterion is then used to accept the sweep.} In the equilibration phase, the maximum displacements and rotation are adjusted such that the acceptance ratio of sweeps stabilizes at $\approx0.25$. (These maximum values are retained without further adjustments in the production phase.)  Configurations are stored every 500 sweeps for further analysis. 

\subsection{Monte Carlo simulations with semi-empirical potentials}\label{sc:pm3}
Once we have the local molecular field, we can, in principle, examine the site with Monte Carlo or molecular dynamics  at a high level of theory. However, even for 59 particles, our initial attempts at using B3LYP/TZV(2d+p) in the Monte Carlo scheme proved intractable, as excessively long times were deemed necessary for adequate convergence. For this reason, in this initial study, we resorted to Monte Carlo simulation of the binding site using the semi-empirical PM3~\cite{stuart:jcc89a,stuart:jcc89b} model in the Gaussian09 package~\cite{g09}.  The simulations are performed with the molecular field and the optimal value of $\alpha$ determined with the coarse potential. The equilibration and production comprises $50 \times 10^3$  and $25\times 10^3$  sweeps respectively. Hydrogen atoms are added on the fly to satisfy valencies of C and N atoms. In the equilibration phase, the acceptance ratio is 0.25. Configurations are saved after every 5 sweeps. 

To compute the free energy change (Eq.~\ref{eq:apmu}) in replacing Zn$^{2+}$ with X$^{2+}(=\mathrm{Co^{2+}, Fe^{2+}, Ni^{2+}})$, we assume that the configurations generated with the PM3 hamiltonian well-represent the configurations that would be produced with the B3LYP/TZV(2d+p) method, and then take 150 equally spaced configurations from the production phase of the PM3-based simulation. For each configuration, the energy change in exchanging Zn$^{2+}$ (Eq.~\ref{eq:apmu}), $\Delta U_1 = U_{1,\rm X^{2+}} - U_{1,\rm Zn^{2+}}$, is calculated at the unrestricted B3LYP/TZV(2d+p) level using GAMESS~\cite{gamess}. Consistent with experiments, we model all metal complexes in their high-spin electronic configuration~\cite{berg:jacs89,berg:ic92}. 

{To the test the effect of suppressing protein restraints, we also compute the free energy of replacing Zn$^{2+}$ with Co$^{2+}$ or Fe$^{2+}$. $50\times 10^3$ sweeps of equilibration and $25\times 10^3$ sweeps of production were performed with the PM3 hamiltonian and
$\alpha = 0.0$. We then take 80 well separated configurations from the production phase to obtain a qualitative estimate of the free energy change with energies obtained at the B3LYP/TZV(2d+p) level.}

For obtaining $\Delta \mu^{\rm ex}(\mathrm{aq})$, the change in the excess free energy in bulk water (Eq.~\ref{eq:delmu}), we borrow from our earlier results based on the primitive quasi-chemical approach \cite{asthagiri:jacs04}. Geometry, thermal corrections, and long-range contributions to the free energy are obtained from the earlier study, but for consistency with the presence work, single point energy calculations are performed at the unrestricted B3LYP/TZV(2d+p) level. 

\section{Results and Discussion\label{sec-results}}
Unless otherwise mentioned, we report the key results of our model extensively only for YTA. Similar agreement is observed for CP1 and TF3. The selective preference for Zn$^{2+}$ over competing Co$^{2+}$, Ni$^{2+}$, and Fe$^{2+}$ has been experimentally measured for CP1~\cite{berg:arbbs97} and serves as a helpful metric to assess the present approach in quantifying thermodynamics of ion binding to protein sites.

\subsection{Validation of the quadratic model}
\begin{figure}
\includegraphics{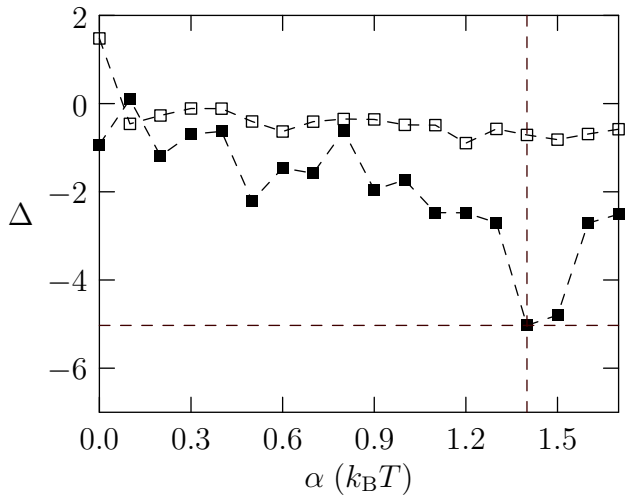}
\caption{$\Delta$ (Eq.~\ref{eq:kl}) as a function of the coupling constant $\alpha$ with (filled) and without (open) site-bulk coupling (see text). The minimum of $\Delta$ is at $\alpha=1.4\, k_{\rm B}T$.}\label{fg:klyta}
\end{figure}

Figure~\ref{fg:klyta} shows $\Delta (\alpha)$ (Eq.~\ref{eq:kl}) between $P_s(\varepsilon)$, the reference binding energy distribution, and $P(\varepsilon;\alpha)$ obtained 
using the effective potential.  $\Delta(\alpha)$ is a minimum at $\alpha=1.4$; this is the value
of $\alpha$ chosen for all further investigations. 
To examine the role of site-bulk interactions in site-site interactions, we set $\mathcal G = \bf 0$ (Eq.~\ref{eq:hmat}); that is, we suppress site-bulk interactions. Physically, this corresponds to 
independent springs on site particles, as opposed to a site of interconnected particles. As 
Figure~\ref{fg:klyta} illustrates, with no site-bulk interactions $\Delta(\alpha)$ is close to zero for all non-zero $\alpha$: the probability distribution for binding energies $P(\varepsilon; \alpha)$ has only a minimal overlap with the target distribution $P_s(\varepsilon)$. Thus, in accordance with experiments indicating the importance of second-shell interactions in tuning metal binding
\cite{Michael:1992p3116,berg:arbbs97} in the zinc-finger protein, the interaction of the site with the bulk material (protein outside the site and the solvent) is important in tuning the binding energy of Zn$^{2+}$ with the site. 

\begin{figure}
\begin{center}
\includegraphics[scale=0.9]{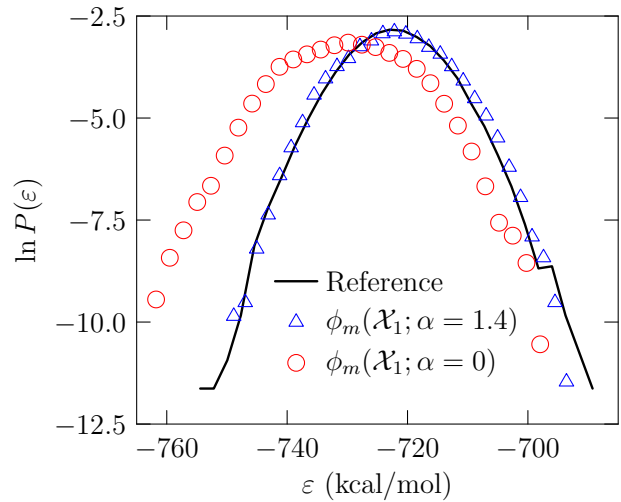}
\end{center}
\caption{Distribution of binding energies of Zn$^{2+}$ with the site from the reference MD simulations (solid curve) and with simulations using the effective potential (Eq.~\ref{eq:pot}) with $\phi_m(\mathcal X_1;\alpha = 0)$ ($\circ$) and with $\phi_m(\mathcal X_1;\alpha =1.4)$ ($\triangle$).  }
\label{fg:zn_yta}	
\end{figure}

Figure~\ref{fg:zn_yta} shows that the site in vacuum ($\alpha = 0$) produces a binding energy distribution that is substantially different from either the MD simulations or with the optimal ($\alpha = 1.4$) site-bulk coupling constant. In particular, in the MD simulation and for the site with the local molecular field, ion-site interactions are shifted to more positive values than for the site in vacuum. Thus, on average, Zn$^{2+}$ is better bound to the cluster in vacuum than it is when bulk protein and solvent effects are considered. Physically this implies that the role of the bulk protein and the solvent is to {\em destabilize} the binding of the ion with the site. Or in other words, {the folded $\beta\beta\alpha$ conformation of the protein is stabilized by the introduction of the metal. Note that experiments suggest that metal coordination is essential for the protein fold and the metal free apo-peptide is unstructured~\cite{Frankel:1987p3085,Michael:1992p3116,berg:arbbs97}.}

\begin{figure}
\begin{center}
\includegraphics[scale=0.9]{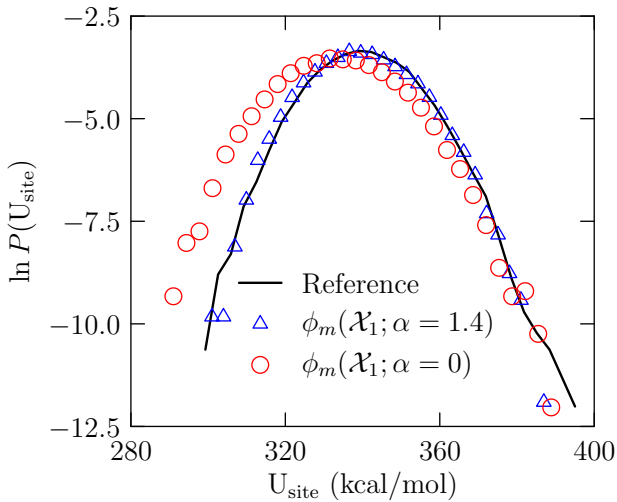}
\end{center}
\caption{Distribution $P({\rm U}_{\rm site})$ of the potential energy ${\rm U}_{\rm site}$ of the site particles
excluding the Zn$^{2+}$ atom. Symbols as in Figure~\ref{fg:zn_yta}. Observes that the site in vacuum is more stable than the site including the bulk protein and solvent response.} \label{fg:sys_lc}
\end{figure}
Figure~\ref{fg:sys_lc} shows that distribution of the potential energy of the site (excluding the Zn$^{2+}$ atom).  The potential energy  ${\rm U}_{\rm site}$ includes contributions from bonding, angles, torsions, improper angles, and non-bonded terms within the site. The site in vacuum is stable by about 6~kcal/mol ($\Delta \langle {\rm U_{site}} \rangle =-6$~kcal/mol between the binding site in vacuum and MD simulations) than the one coupled to the protein and solvent, again emphasizing the role of the metal
in stabilizing the protein fold.  Note that the coupling constant, $\alpha$, was chosen to match the binding energy profile, $P(\varepsilon)$ of the Zn$^{2+}$ ion. The agreement in the distribution, $P({\rm U}_{\rm site})$, of the binding site energetics is an independent verification of our model. Further, this observation suggests that the ENM can be successfully parametrized with respect to either ion binding energetics or site energetics.

As we have argued before \cite{dixit:bj09,dixit:bj11}, the conformations that correspond to 
low ${\rm U}_{\rm site}$, that is conformations for which the site is less strained energetically, 
are also the ones for which the ion-site interactions are unfavorable for the ion: the less strained conformations of the site correspond to the high energy tail of the ion-site distribution. As Eq.~\ref{eq:pdt} emphasizes, these conformations are also 
the ones that sensitively influence the excess chemical potential of the ion in the site. In this context it is important to note that the local molecular field is able to capture both the low energy wing of $P({\rm U}_{\rm site})$ (Figure~\ref{fg:sys_lc}) and the high-energy wing of $P(\varepsilon)$ (Figure~\ref{fg:zn_yta}).

\subsection{Structural characterization of the binding site}
In Table~\ref{table-geom} we summarize the various structural parameters that are relevant to the geometry of the residues around the Zn$^{2+}$ atom. We find that Zn$^{2+}$ is able to maintain a  tetra-coordinate state throughout the course of the simulation: the sulphur and nitrogen atoms from the cystine and histidine residues, respectively, are always coordinated with the metal. 

The geometry of the binding site in vacuum ($\alpha = 0$) and with the optimal site-bulk coupling in the effective potential (Eq.~\ref{eq:pot}) agree reasonably well with MD simulations. The  $\angle$N-Zn$^{2+}$-N angle is somewhat more expanded in the MD simulations and the $\angle$S-Zn$^{2+}$-S is somewhat more compressed relative to the site with either $\alpha = 0$ or $\alpha = 1.4$. But the fluctuations are large and it is unclear if these differences are significant.  This observation suggests that in tuning the molecular molecular field, energies may prove more sensitive than structural parameters. 

\subsection{Investigation of the protein restraints}
It is known that the folding of the zinc finger peptide is coupled with metal binding~\cite{Frankel:1987p3085,Michael:1992p3116,berg:arbbs97}. Thus, residues which are crucial in maintaining the binding site are also expected to be important in the folding of the peptide. It has been suggested that the $\beta$-sheet hosting the two cysteine residues is formed in the absence of Zn$^{2+}$ and the addition of the metal induces the folding of the $\alpha$-helix~\cite{berg:bj93,takeuchi:bba98,hanas:2005}. The coordination of Zn$^{2+}$ with the histidine residues is thought to be essential towards folding of the $\alpha$-helix and formation of the hydrophobic core. We next consider how analyzing the protein field can provide insights into some of these features.
\begin{table}[h]
\begin{center}
\begin{tabular}{c|ccc}
& MD&MC $\alpha=0$&MC $\alpha=1.4$  \\ \hline
r$_{\rm S-Zn^{2+}}$ &	2.17 $\pm$ 0.03 & 2.16 $\pm$ 0.03  & 2.15 $\pm$ 0.03  \\
r$_{\rm N-Zn^{2+}}$&	2.08 $\pm$ 0.04 &  2.1 $\pm$ 0.05 & 2.07 $\pm$ 0.03  \\
$\angle$S-Zn$^{2+}$-S&112.0 $\pm$ 4.5 & 120.2 $\pm$ 5.9 & 121.9 $\pm$ 4.2  \\
$\angle$N-Zn$^{2+}$-N&104.6 $\pm$ 4.6 & 102.2 $\pm$ 4.6 & 100.5 $\pm$ 3.2  \\
\end{tabular}
\caption{Bond lengths and angles characterizing the geometry of the site. All angles are in degrees and distances in {\AA}ngstroms. MD, molecular dynamics; $\alpha = 0$ (vacuum) and $\alpha = 1.4\, k_{\rm B}T$, correspond to Monte Carlo simulations with the effective potential (Eq.~\ref{eq:pot}). Standard deviations of the quantities are noted.\label{table-geom}}
\end{center}
\end{table}

The restraints imposed by the protein on the binding site, $\phi_{\rm m}(\mathcal X_1;\alpha)$ in Eq.~\ref{eq:pot}, are collective in nature and cannot be expressed as a sum of independent restraints over individual atoms of the binding site. The quadratic approximation decomposes the restraints as a sum over mutually orthogonal collective motions of the binding site atoms. These collective motions or vibrational modes are along the eigenvectors of the positive definite matrix, $\mathcal H_{\rm site}$ in Eq.~\ref{eq:pot}, describing the protein field, $\phi_{\rm m}(\mathcal X_1,;\alpha)$. 
\begin{figure*}
\includegraphics[scale=0.95]{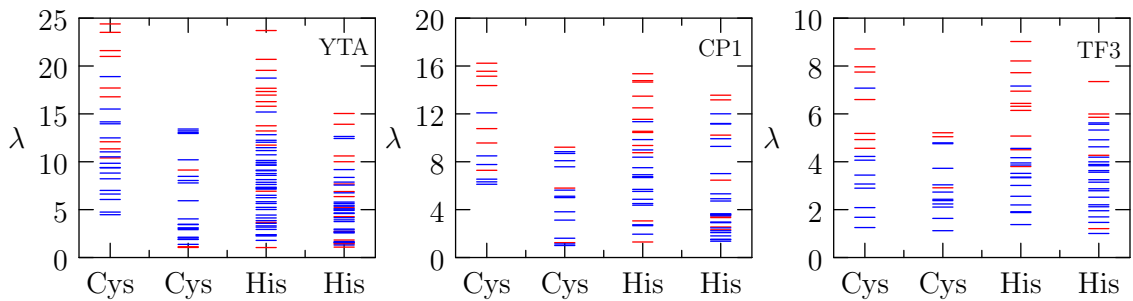}
\caption{ Vibrational modes for YTA, CP1, and TF3  arranged according to the residue of the binding site which contributes most to its magnitude. 
The  eigenvalue corresponding to each mode in {$k_{\rm B}T/{\rm \AA}^2$} is plotted for each of the four metal-binding residues. Red lines denote concentrated modes while blue lines denote distributed ones. Observe that the strongest vibrational modes are concentrated on the first cysteine and the first histidine of the CCHH cluster comprising the binding site. }\label{sp:yta}
\end{figure*}

We classify the vibrational modes as either concentrated on a single residue or distributed on more than one residue.  Since the eigenvector has contributions from each of the atom comprising the binding site, we consider the contribution to the norm of the eigenvector from each residue of the binding site to evaluate the behavior of the modes. We call the mode concentrated if the majority of the contribution to the norm is found on one residue (We choose a threshold of 0.8 for delineating concentrated modes from all other modes. The qualitative insights below are independent of this threshold.) Following this procedure for all the modes, Fig.~\ref{fg:sp_yta} shows that the most important protein restraints for three different zinc finger domains are in fact concentrated on the first cysteine and the first histidine residue of the CCHH cluster comprising the binding site. This implies that the first C and the first H play an important role in maintaining the architecture of the site.

Nomura and Sugiura~\cite{sugiura:ic02} conducted experiments on the Sp$_1$ zinc finger domain where they mutated one residue at a time in the CCHH binding site to a glycine residue, eliminating the propensity of that particular residue to coordinate with the Zn$^{2+}$ center. The authors showed, through circular dichroism  measurements, that the CCHG mutant of the protein is able to form an $\alpha$-helix and a $\beta$-sheet structure similar to that of the wild type protein, while no helical structure is formed with the CCGH mutant, implying the importance of the first histidine in maintaining the architecture of the binding site. Our analysis of concentrated modes is consistent with this experimental observation of the importance of the first histidine in the CCHH cluster.

Since the $\beta$-sheet hosting the first cysteine is formed prior to the introduction of the metal, analysis of the modes in the presence of the metal is not sufficient to infer the consequences of mutations in the cysteine residues. Our analysis does suggest that the Zn$^{2+}$ will be loosely held in the GCHH finger and experience larger fluctuations compared to the CGHH finger. Spectroscopic experiments to probe the dynamics of the binding site can be helpful in evaluating this suggestion.

Fig.~\ref{fg:sp_yta} shows that the eigenvalues $\lambda$ defining the strength of the restraints on the binding site are larger for the designed peptides CP1 and YTA than for TF3: deflections of the binding site from the equilibrium structure require more energy for the designed peptides than for the natural TF3 peptide. {Thus we expect CP1 and YTA to experience smaller deviations from the equilibrium structure as compared to TF3. Since metal coordination has a stabilizing effect on the protein fold and since deformations from the equilibrium structure
will also tend to destabilize Zn$^{2+}$ binding to the protein, we can induce that CP1 and YTA will experience a higher stabilization due to introduction of the metal as compared to TF3.} It is interesting to note that CP1 was designed by aligning 131 natural zinc finger sequences and is found to have a greater affinity for zinc than the corresponding natural sequences \cite{Krizek:jacs91}. The predicted greater stabilization
of Zn$^{2+}$-bound CP1 is in accordance with this experimental observation.   

{Sequence specific effects on stabilization can also be inferred. In CP1 and YTA only two residues interleave the cysteine residues in contrast to four residues in the natural peptide.  This shorter turn between the cysteine residues likely underlies the observed differences in binding energetics, but a thorough exploration of  how such sequence dependences influence binding is left for future investigations. The analysis of concentrated modes developed here may prove helpful in this regard.}

\subsection{Selectivity of the site for Zn$^{2+}$ over competing transition metals}
Table~\ref{table-dmu} summarizes the {predicted values for the} free energy difference between Zn$^{2+}$ and competing divalent ions for the consensus peptide CP1. The order of selectivity $\mathrm {Co^{2+} < Ni^{2+}<Fe^{2+}}$ relative to Zn$^{2+}$ is consistent with experiments~\cite{berg:arbbs97}. The magnitude of selectivity for Fe$^{2+}$ is also in good agreement with experiments.
\begin{table}[h]
\begin{tabular}{ccrrrc}
\hline
Ion& $\alpha$ & $\Delta \mu^{\rm ex}$(S)& $\Delta \mu^{\rm ex}$(aq)& $\Delta \mu^{\rm ex}$ (Calc.) &$\Delta \mu^{\rm ex}$ (Expt.)\\
\hline
Co$^{2+}$& 1.4 & 6.3  &5.2  & 1.1 &5.4 \\
     & 0.0 & 6.9 & 5.2 & 1.7 & 5.4  \\
Fe$^{2+}$& 1.4 & 29.2  &20.4  & 8.4 &7.8 \\
& 0.0 & 16.9 & 20.4 & $-3.5$ & 7.8 \\
Ni$^{2+}$ & 1.4 & $-7.7$  & $-12.6$  & 4.3 &7.4\\ \hline
\end{tabular}
\caption{Free energy to replace Zn$^{2+}$ Co$^{2+}$, Fe$^{2+}$, and Ni$^{2+}$  (Eq.~\ref{eq:delmu}) in the CP1 peptide. All values are
in kcal/mol.  $\Delta \mu^{\rm ex}(\mathrm{S})$ is calculated using Eq.~\ref{eq:apmu}.  Calculations for a cluster in vacuum ($\alpha = 0$) and for the cluster in a  molecular field ($\alpha = 1.4\, k_{\rm B}T $) are reported for Co$^{2+}$ and Fe$^{2+}$.  $\Delta \mu^{\rm ex}$(aq) was obtained using primitive quasi-chemical theory \cite{asthagiri:jacs04}. (Please refer to Sec.~\ref{sc:pm3} for details.) Experimental estimates from Ref.~\cite{berg:arbbs97} are provided for comparison. The calculated selectivities for Co$^{2+}$, Fe$^{2+}$, and Ni$^{2+}$, respectively, in the YTA peptide  are $1.3$, $8.1$, and $6.7$, similar to the estimate for CP1. We are not aware of experiments characterizing the selectivity in YTA.\label{table-dmu}}
\end{table}

We first consider limitations in calculating $\Delta\mu^{\rm ex}$(aq) in affecting the predicted selectivity, 
$\Delta\mu^{\rm ex}$ (Eq.~\ref{eq:delmu}). The aqueous component of the selectivity free energy, $\Delta\mu^{\rm ex}$(aq), is only about 1\% of the absolute hydration free energy of the metal ion \cite{asthagiri:jacs04}; thus small errors in the absolute hydration free energy can be amplified in taking differences. In the present study,  we have used a primitive quasichemical estimate \cite{asthagiri:jacs04} for  $\Delta \mu^{\rm ex}$(aq): the free energy of forming a metal ion-water cluster in vacuum is combined with a continuum dielectric correction for the presence of the bulk material. Ignoring the role of the bulk material in forming the metal-water cluster \cite{merchant:jcp09} and potential limitations in a continuum dielectric description of water beyond the first shell for highly charged metals \cite{lrp:ionsjcp03,rode:znjcp} are important limitations in our calculation of $\Delta\mu^{\rm ex}$(aq).

{Limitations in calculating $\Delta\mu^{\rm ex}(\rm S)$ is the other factor in affecting the predicted selectivity. Examining the distribution of $\Delta U_1$ (Eq.~\ref{eq:apmu}), the difference in the binding energy of metal ${\rm X}^{\rm 2+}$ with the binding site relative to the corresponding value for Zn$^{2+}$,  shows that $\Delta U_1$ is narrowly distributed about the mean value $\langle \Delta U_1\rangle$ (Table~\ref{table-cumu}), although the distribution of the binding energy of the metal with the cluster itself is broad (Fig.~\ref{fg:zn_yta}). Thus the
calculated free energy differences in the presence of the field are expected to be reasonably well-converged.} 

{Table~\ref{table-dmu} shows  that the protein field plays a decisive role in selectivity. The protein restraints limit the phase space distribution of the binding site to configurations which bind to Zn$^{2+}$ strongly. Removing these restraints allows the binding site to sample configurations which bind favorably to Fe$^{2+}$ as well.  We find that the free energy for replacing Zn$^{2+}$ with Fe$^{2+}$ without the protein field ($\alpha = 0$) predicts greater stabilization of the zinc-finger in the presence of Fe$^{2+}$, in complete disagreement with experiments.  The role of the protein in replacing Zn$^{2+}$ with Co$^{2+}$ is predicted
to be negligible. Examining the  distribution of $\Delta U_1$ shows that the protein tends to limit energy fluctuations ($C_2$ term, Table~\ref{table-cumu}) and the effect is more pronounced for metal ions that interact strongly (relative to Zn$^{2+}$) with the zinc-binding residues, as the comparison of Fe$^{2+}$ and Co$^{2+}$ illustrates (cf. $C_1$ and $C_2$ Table~\ref{table-cumu}). 
Taken together, these observations show that while the local metal-residue interaction is important in selectivity, as was already inferred in the early studies on zinc-fingers\cite{berg:jacs89}, the role of the protein cannot be ignored, a result that is in accordance with experimental investigations comparing two different zinc-binding proteins \cite{weiss:biochem04}. }

{An important limitation in the present study is the small number of configurations sampled at  the semi-empirical level. The small magnitude of $C_2$ relative to $C_1$ suggests that this limitation may not be severe in the case when the
protein field is present. In any case, with a more efficient implementation of the present approach, this limitation can always be overcome. 
Regardless of these computational limitations, the important physical point we wish to emphasize is that accounting for the protein field is necessary for a broader characterization of the thermodynamics of metal-ion binding in this system. Thus, for example, the protein
field will be necessary for a proper description of excess entropies and excess energies, quantities that demand a 
proper description of  the role of the medium on the local ion-residue cluster \cite{dixit:bj11}.}
\begin{table}[h]
\begin{center}
\begin{tabular}{ccrrrr}
\hline
Ion& $\alpha$  & $\Delta \mu^{\rm ex}$(S) & $C_1$ & $C_2$ & $C_1 - C_2$ \\
\hline
Co$^{2+}$&1.4 & 6.3  & 6.8 & 0.5 & 6.3 \\
                   & 0.0 & 6.9 & 7.6 & 0.9 & 6.7 \\
Fe$^{2+}$ &1.4 & 29.2  & 29.2  & 0.5  & 28.7 \\
                   & 0.0 & 16.9 & 29.8 & 37.0 & $-7.2$ \\
Ni$^{2+}$ &1.4 & $-7.7$  & $-7.1$ & 1.3 & $-8.4$ \\ \hline
\end{tabular}
\caption{Comparing the exponential average (Eq.~\ref{eq:delmu}) with the Gaussian model $\Delta \mu^{\rm ex}(\mathrm{S}) = C_1 - C_2$, where $C_1 = \langle \Delta U_1\rangle$ and $C_2 = \beta/2 \langle (\Delta U_1 - C_1)^2 \rangle$. $C_1$ and $2 C_2/\beta $ are the first (mean) and second (variance) moments of the distribution of $\Delta U_1$ values \cite{Hummer:jpca98}. Calculations for a cluster in vacuum ($\alpha = 0$) and for the cluster in a  molecular field ($\alpha = 1.4\, k_{\rm B}T$) are reported for Co$^{2+}$ and Fe$^{2+}$. All values in kcal/mol.} \label{table-cumu}
\end{center}
\end{table}

Changes in the equilibrium geometry of the protein can be expected with ion exchange owing to change in the ion size, which, within the quadratic approximation, will change the harmonic restraints on the binding site and thus the estimate of the selectivity free energy. Here $\Delta\mu^{\rm ex}$(S) was calculated with the assumption that the molecular field imposed by the protein medium and the solvent is independent of the bound ion, $\Delta \phi \approx 0$. For the given coordination environment, 
the radii for metals considered here are expected to be similar and so assuming $\Delta\phi \approx 0$ appears reasonable. This assumption will certainly not hold when a metal associates with the binding site in a geometry substantially different from the one for Zn$^{2+}$ as may happen for Hg$^{2+}$~\cite{lim:jacs09}.

\section{Concluding Discussions\label{sec-disc}}

The architecture and conformation of the metal binding site in metalloproteins is conditioned by energetic interactions of the metal cation and the binding site on the one hand and the interaction of the binding site with the protein and solvent media outside the binding site on the other. In the case of the zinc finger peptide, reflecting the fact that the peptide is 
unfolded in the absence of the metal,  calculations show that including the interaction of the site with the 
protein material outside the binding site results in a more {\em destabilized} metal-binding site complex in
comparison to an analogous metal-binding site complex in vacuum. 

The protein outside the metal-residue cluster imposes a field on the cluster.  By treating the bulk protein as an elastic medium, 
we describe the molecular field by quadratic model. This simplifies the problem of tackling metal protein interactions by greatly  reducing the number of degrees of freedom and allows us to study the effect of the protein medium on the binding site in a transparent fashion. Decomposing the protein restraints on the binding site over mutually orthogonal collective 
motions of the binding site particles shows that the protein restraints are stronger for designed zinc-finger peptides CP1 and YTA 
relative to the natural zinc finger TF3.  Further, for all three zinc finger peptides, we find that the first cysteine and first histidine in the CCHH binding site  are more tightly held by the protein as compared to the other two residues of the binding site. Our results are in accordance with the importance of the first histidine observed on the basis of metal-induced folding experiments with a CCGH mutant binding site. We also predict that were a peptide with the GCHH binding motif to fold in the presence of {Zn$^{2+}$,} 
that metal-residue cluster is expected to experience large fluctuations relative to the CCHH binding motif. 

Approximating the solvent and the protein response by a quadratic term implies that we have neglected specific interactions of solvent molecules on the dynamics of the binding site. Care would be required when the present model is applied to ion binding sites where water molecules play a crucial role. The molecular field approach developed here requires the equilibrium structure of the protein in order to estimate the response of the bulk protein due to the binding site. If this response is not sensitively dependent on the bound ion, the molecular field obtained on the basis of a protein structure for one ion can be used to predict the free energy change in replacing that ion with a competing one. Such calculations on the zinc-finger peptide give reasonable estimates of the free energy of replacing Zn$^{2+}$ with Co$^{2+}$, Fe$^{2+}$, and Ni$^{2+}$.

\section*{Acknowledgments}
PD thanks Christian Micheletti for introducing him to elastic network models. We thank Christian Micheletti and Safir Merchant for helpful comments on the manuscript  This research used resources of the National Energy Research Scientific Computing Center, which is supported by the Office of Science of the U.S. Department of Energy under Contract No. DE- AC02-05CH11231.

\section*{Supplementary Information}
Complete reference to Frisch M. J. et al. ~\cite{g09}, CHARMM \cite{charmm:jpcb98}, and additional data for CP1 and TF3 peptides

\newpage

\begin{mcitethebibliography}{62}
\providecommand*\natexlab[1]{#1}
\providecommand*\mciteSetBstSublistMode[1]{}
\providecommand*\mciteSetBstMaxWidthForm[2]{}
\providecommand*\mciteBstWouldAddEndPuncttrue
  {\def\EndOfBibitem{\unskip.}}
\providecommand*\mciteBstWouldAddEndPunctfalse
  {\let\EndOfBibitem\relax}
\providecommand*\mciteSetBstMidEndSepPunct[3]{}
\providecommand*\mciteSetBstSublistLabelBeginEnd[3]{}
\providecommand*\EndOfBibitem{}
\mciteSetBstSublistMode{f}
\mciteSetBstMaxWidthForm{subitem}{(\alph{mcitesubitemcount})}
\mciteSetBstSublistLabelBeginEnd
  {\mcitemaxwidthsubitemform\space}
  {\relax}
  {\relax}

\bibitem[Bren et~al.(2004)Bren, Pecoraro, and Gray]{gray:ic04}
Bren,~K.~L.; Pecoraro,~V.~L.; Gray,~H.~B. \emph{Inorg. Chem.} \textbf{2004},
  \emph{43}, 7894--7896\relax
\mciteBstWouldAddEndPuncttrue
\mciteSetBstMidEndSepPunct{\mcitedefaultmidpunct}
{\mcitedefaultendpunct}{\mcitedefaultseppunct}\relax
\EndOfBibitem
\bibitem[Lippard and Berg(1994)Lippard, and Berg]{bergbook}
Lippard,~S.~J.; Berg,~J.~M. \emph{Principles of bioinorganic chemistry};
  University Science Books: Mill Valley, CA, 1994\relax
\mciteBstWouldAddEndPuncttrue
\mciteSetBstMidEndSepPunct{\mcitedefaultmidpunct}
{\mcitedefaultendpunct}{\mcitedefaultseppunct}\relax
\EndOfBibitem
\bibitem[Millhauser(2004)]{Millhauser:acc04}
Millhauser,~G.~L. \emph{Acc. Chem. Res.} \textbf{2004}, \emph{37}, 79--85\relax
\mciteBstWouldAddEndPuncttrue
\mciteSetBstMidEndSepPunct{\mcitedefaultmidpunct}
{\mcitedefaultendpunct}{\mcitedefaultseppunct}\relax
\EndOfBibitem
\bibitem[Wilson et~al.(2004)Wilson, Apiyo, and Wittung-Stafshede]{Wilson:qrb04}
Wilson,~C.~J.; Apiyo,~D.; Wittung-Stafshede,~P. \emph{Q. Rev. Biophys.}
  \textbf{2004}, \emph{37}, 285--314\relax
\mciteBstWouldAddEndPuncttrue
\mciteSetBstMidEndSepPunct{\mcitedefaultmidpunct}
{\mcitedefaultendpunct}{\mcitedefaultseppunct}\relax
\EndOfBibitem
\bibitem[Adlard and Bush(2006)Adlard, and Bush]{Adlard:jad06}
Adlard,~P.~A.; Bush,~A.~I. \emph{J. Alzheimers Dis.} \textbf{2006}, \emph{10},
  145--163\relax
\mciteBstWouldAddEndPuncttrue
\mciteSetBstMidEndSepPunct{\mcitedefaultmidpunct}
{\mcitedefaultendpunct}{\mcitedefaultseppunct}\relax
\EndOfBibitem
\bibitem[Siegbahn and Blomberg(1999)Siegbahn, and Blomberg]{siegbahn:rev99}
Siegbahn,~P. E.~M.; Blomberg,~M. R.~A. \emph{Ann. Rev. Phys. Chem.}
  \textbf{1999}, \emph{50}, 221--249\relax
\mciteBstWouldAddEndPuncttrue
\mciteSetBstMidEndSepPunct{\mcitedefaultmidpunct}
{\mcitedefaultendpunct}{\mcitedefaultseppunct}\relax
\EndOfBibitem
\bibitem[Ryde(2003)]{ryde:cocb03}
Ryde,~U. \emph{Curr. Opin. Chem. Biol.} \textbf{2003}, \emph{7}, 136--142\relax
\mciteBstWouldAddEndPuncttrue
\mciteSetBstMidEndSepPunct{\mcitedefaultmidpunct}
{\mcitedefaultendpunct}{\mcitedefaultseppunct}\relax
\EndOfBibitem
\bibitem[Lin and Truhlar(2007)Lin, and Truhlar]{truhlar:tca07}
Lin,~H.; Truhlar,~D.~G. \emph{Theor. Chem. Acc.} \textbf{2007}, \emph{117},
  185--199\relax
\mciteBstWouldAddEndPuncttrue
\mciteSetBstMidEndSepPunct{\mcitedefaultmidpunct}
{\mcitedefaultendpunct}{\mcitedefaultseppunct}\relax
\EndOfBibitem
\bibitem[Senn and Thiel(2009)Senn, and Thiel]{thiel:angew09}
Senn,~H.~M.; Thiel,~W. \emph{Angew. Chem. Intl. Ed.} \textbf{2009}, \emph{48},
  1198--1229\relax
\mciteBstWouldAddEndPuncttrue
\mciteSetBstMidEndSepPunct{\mcitedefaultmidpunct}
{\mcitedefaultendpunct}{\mcitedefaultseppunct}\relax
\EndOfBibitem
\bibitem[Pratt and Ashbaugh(2003)Pratt, and Ashbaugh]{lrp:hspre}
Pratt,~L.~R.; Ashbaugh,~H.~S. \emph{Phys. Rev. E} \textbf{2003}, \emph{68},
  021505\relax
\mciteBstWouldAddEndPuncttrue
\mciteSetBstMidEndSepPunct{\mcitedefaultmidpunct}
{\mcitedefaultendpunct}{\mcitedefaultseppunct}\relax
\EndOfBibitem
\bibitem[Zheng and Brooks(2006)Zheng, and Brooks]{brooks:bj06}
Zheng,~W.; Brooks,~B.~R. \emph{Biophys. J.} \textbf{2006}, \emph{88},
  3109--3117\relax
\mciteBstWouldAddEndPuncttrue
\mciteSetBstMidEndSepPunct{\mcitedefaultmidpunct}
{\mcitedefaultendpunct}{\mcitedefaultseppunct}\relax
\EndOfBibitem
\bibitem[Zheng et~al.(2007)Zheng, Liao, Brooks, and Doniach]{brooks:ps07}
Zheng,~W.; Liao,~J.~C.; Brooks,~B.~R.; Doniach,~S. \emph{Proteins: Struc. Func.
  Bioinfo.} \textbf{2007}, \emph{67}, 886--896\relax
\mciteBstWouldAddEndPuncttrue
\mciteSetBstMidEndSepPunct{\mcitedefaultmidpunct}
{\mcitedefaultendpunct}{\mcitedefaultseppunct}\relax
\EndOfBibitem
\bibitem[Woodcock et~al.(2008)Woodcock, Zheng, Ghysels, Shao, Kong, and
  Brooks]{brooks:jcp08}
Woodcock,~H.~L.; Zheng,~W.; Ghysels,~A.; Shao,~Y.; Kong,~J.; Brooks,~B.~R.
  \emph{J. Chem. Phys.} \textbf{2008}, \emph{129}, 214109\relax
\mciteBstWouldAddEndPuncttrue
\mciteSetBstMidEndSepPunct{\mcitedefaultmidpunct}
{\mcitedefaultendpunct}{\mcitedefaultseppunct}\relax
\EndOfBibitem
\bibitem[Atilgan et~al.(2001)Atilgan, Durell, Jernigan, Demirel, Keskin, and
  Bahar]{bahar:bj01}
Atilgan,~A.~R.; Durell,~S.~R.; Jernigan,~R.~L.; Demirel,~M.~C.; Keskin,~O.;
  Bahar,~I. \emph{Biophys. J.} \textbf{2001}, \emph{80}, 505--515\relax
\mciteBstWouldAddEndPuncttrue
\mciteSetBstMidEndSepPunct{\mcitedefaultmidpunct}
{\mcitedefaultendpunct}{\mcitedefaultseppunct}\relax
\EndOfBibitem
\bibitem[Bahar and Reader(2005)Bahar, and Reader]{bahar:cosb05}
Bahar,~I.; Reader,~A.~J. \emph{Curr. Opin. Struc. Biol.} \textbf{2005},
  \emph{15}, 586--592\relax
\mciteBstWouldAddEndPuncttrue
\mciteSetBstMidEndSepPunct{\mcitedefaultmidpunct}
{\mcitedefaultendpunct}{\mcitedefaultseppunct}\relax
\EndOfBibitem
\bibitem[Micheletti et~al.(2004)Micheletti, Carloni, and
  Maritan]{micheletti:ps04}
Micheletti,~C.; Carloni,~P.; Maritan,~A. \emph{Proteins: Struc. Func. Bioinfo.}
  \textbf{2004}, \emph{55}, 635--645\relax
\mciteBstWouldAddEndPuncttrue
\mciteSetBstMidEndSepPunct{\mcitedefaultmidpunct}
{\mcitedefaultendpunct}{\mcitedefaultseppunct}\relax
\EndOfBibitem
\bibitem[Pontiggia et~al.(2007)Pontiggia, Colombo, Micheletti, and
  Orland]{micheletti:prl07}
Pontiggia,~F.; Colombo,~G.; Micheletti,~C.; Orland,~H. \emph{Phys. Rev. Lett.}
  \textbf{2007}, \emph{98}, 048102\relax
\mciteBstWouldAddEndPuncttrue
\mciteSetBstMidEndSepPunct{\mcitedefaultmidpunct}
{\mcitedefaultendpunct}{\mcitedefaultseppunct}\relax
\EndOfBibitem
\bibitem[Capozzi et~al.(2007)Capozzi, Luchinat, Micheletti, and
  Pontiggia]{micheletti:jpr07}
Capozzi,~F.; Luchinat,~C.; Micheletti,~C.; Pontiggia,~F. \emph{J. Proteome
  Res.} \textbf{2007}, \emph{6}, 4245--4255\relax
\mciteBstWouldAddEndPuncttrue
\mciteSetBstMidEndSepPunct{\mcitedefaultmidpunct}
{\mcitedefaultendpunct}{\mcitedefaultseppunct}\relax
\EndOfBibitem
\bibitem[Zen et~al.(2009)Zen, Carnevale, Lesk, and Micheletti]{micheletti:ps09}
Zen,~A.; Carnevale,~V.; Lesk,~A.~M.; Micheletti,~C. \emph{Protein Science}
  \textbf{2009}, \emph{17}, 918--929\relax
\mciteBstWouldAddEndPuncttrue
\mciteSetBstMidEndSepPunct{\mcitedefaultmidpunct}
{\mcitedefaultendpunct}{\mcitedefaultseppunct}\relax
\EndOfBibitem
\bibitem[Asthagiri et~al.(2006)Asthagiri, Pratt, and
  Paulaitis]{asthagiri:jcp06}
Asthagiri,~D.; Pratt,~L.~R.; Paulaitis,~M.~E. \emph{J. Chem. Phys.}
  \textbf{2006}, \emph{125}, 24701\relax
\mciteBstWouldAddEndPuncttrue
\mciteSetBstMidEndSepPunct{\mcitedefaultmidpunct}
{\mcitedefaultendpunct}{\mcitedefaultseppunct}\relax
\EndOfBibitem
\bibitem[Pratt and Asthagiri(2007)Pratt, and Asthagiri]{lrp:cpms}
Pratt,~L.~R.; Asthagiri,~D. In \emph{Free energy calculations: {Theory} and
  applications in chemistry and biology}; Chipot,~C., Pohorille,~A., Eds.;
  Springer series in {Chemical Physics}; Springer, 2007; Vol.~86; Chapter 9, pp
  323--351\relax
\mciteBstWouldAddEndPuncttrue
\mciteSetBstMidEndSepPunct{\mcitedefaultmidpunct}
{\mcitedefaultendpunct}{\mcitedefaultseppunct}\relax
\EndOfBibitem
\bibitem[Dixit et~al.(2009)Dixit, Merchant, and Asthagiri]{dixit:bj09}
Dixit,~P.~D.; Merchant,~S.; Asthagiri,~D. \emph{Biophys. J.} \textbf{2009},
  \emph{96}, 2138--2145\relax
\mciteBstWouldAddEndPuncttrue
\mciteSetBstMidEndSepPunct{\mcitedefaultmidpunct}
{\mcitedefaultendpunct}{\mcitedefaultseppunct}\relax
\EndOfBibitem
\bibitem[Dixit and Asthagiri(2011)Dixit, and Asthagiri]{dixit:bj11}
Dixit,~P.~D.; Asthagiri,~D. \emph{Biophys. J.} \textbf{2011}, \emph{100},
  1542--1549\relax
\mciteBstWouldAddEndPuncttrue
\mciteSetBstMidEndSepPunct{\mcitedefaultmidpunct}
{\mcitedefaultendpunct}{\mcitedefaultseppunct}\relax
\EndOfBibitem
\bibitem[Klug and Schwabe(1995)Klug, and Schwabe]{Klug:1995p3038}
Klug,~A.; Schwabe,~J.~W. \emph{{FASEB} J.} \textbf{1995}, \emph{9},
  597--604\relax
\mciteBstWouldAddEndPuncttrue
\mciteSetBstMidEndSepPunct{\mcitedefaultmidpunct}
{\mcitedefaultendpunct}{\mcitedefaultseppunct}\relax
\EndOfBibitem
\bibitem[Berg and Shi(1996)Berg, and Shi]{berg:sc96}
Berg,~J.; Shi,~Y. \emph{Science} \textbf{1996}, \emph{271}, 1081--1085\relax
\mciteBstWouldAddEndPuncttrue
\mciteSetBstMidEndSepPunct{\mcitedefaultmidpunct}
{\mcitedefaultendpunct}{\mcitedefaultseppunct}\relax
\EndOfBibitem
\bibitem[Wolfe et~al.(2000)Wolfe, Nekludova, and Pabo]{Wolfe:2000p3044}
Wolfe,~S.~A.; Nekludova,~L.; Pabo,~C.~O. \emph{Ann. Rev. Biophys. Biomol.
  Struc.} \textbf{2000}, \emph{29}, 183--212\relax
\mciteBstWouldAddEndPuncttrue
\mciteSetBstMidEndSepPunct{\mcitedefaultmidpunct}
{\mcitedefaultendpunct}{\mcitedefaultseppunct}\relax
\EndOfBibitem
\bibitem[Frankel et~al.(1987)Frankel, Berg, and Pabo]{Frankel:1987p3085}
Frankel,~A.~D.; Berg,~J.~M.; Pabo,~C.~O. \emph{Proc. Natl. Acad. Sc. USA}
  \textbf{1987}, \emph{84}, 4841--4845\relax
\mciteBstWouldAddEndPuncttrue
\mciteSetBstMidEndSepPunct{\mcitedefaultmidpunct}
{\mcitedefaultendpunct}{\mcitedefaultseppunct}\relax
\EndOfBibitem
\bibitem[Michael et~al.(1992)Michael, Kilfoil, Schmidt, Amann, and
  Berg]{Michael:1992p3116}
Michael,~S.~F.; Kilfoil,~V.~J.; Schmidt,~M.~H.; Amann,~B.~T.; Berg,~J.~M.
  \emph{Proc. Natl. Acad. Sc. USA} \textbf{1992}, \emph{89}, 4796--800\relax
\mciteBstWouldAddEndPuncttrue
\mciteSetBstMidEndSepPunct{\mcitedefaultmidpunct}
{\mcitedefaultendpunct}{\mcitedefaultseppunct}\relax
\EndOfBibitem
\bibitem[Berg and Godwin(1997)Berg, and Godwin]{berg:arbbs97}
Berg,~J.; Godwin,~H.~A. \emph{Ann. Rev. Biophys. Biomol. Struc.} \textbf{1997},
  \emph{26}, 357--371\relax
\mciteBstWouldAddEndPuncttrue
\mciteSetBstMidEndSepPunct{\mcitedefaultmidpunct}
{\mcitedefaultendpunct}{\mcitedefaultseppunct}\relax
\EndOfBibitem
\bibitem[Reuter et~al.(2000)Reuter, Dejegere, Maigret, and
  Karplus]{karplus:jpca00}
Reuter,~N.; Dejegere,~A.; Maigret,~B.; Karplus,~M. \emph{J. Phys. Chem. A}
  \textbf{2000}, \emph{104}, 1720--1735\relax
\mciteBstWouldAddEndPuncttrue
\mciteSetBstMidEndSepPunct{\mcitedefaultmidpunct}
{\mcitedefaultendpunct}{\mcitedefaultseppunct}\relax
\EndOfBibitem
\bibitem[Ming and Wall(2005)Ming, and Wall]{ming:prl05}
Ming,~D.; Wall,~M.~E. \emph{Phys. Rev. Lett.} \textbf{2005}, \emph{95},
  198103\relax
\mciteBstWouldAddEndPuncttrue
\mciteSetBstMidEndSepPunct{\mcitedefaultmidpunct}
{\mcitedefaultendpunct}{\mcitedefaultseppunct}\relax
\EndOfBibitem
\bibitem[Beck et~al.(2006)Beck, Paulaitis, and Pratt]{lrp:book}
Beck,~T.~L.; Paulaitis,~M.~E.; Pratt,~L.~R. \emph{The potential distribution
  theorem and models of molecular solutions}; Cambridge University Press,
  2006\relax
\mciteBstWouldAddEndPuncttrue
\mciteSetBstMidEndSepPunct{\mcitedefaultmidpunct}
{\mcitedefaultendpunct}{\mcitedefaultseppunct}\relax
\EndOfBibitem
\bibitem[Widom(1982)]{widom:jpc82}
Widom,~B. \emph{J. Phys. Chem.} \textbf{1982}, \emph{86}, 869--872\relax
\mciteBstWouldAddEndPuncttrue
\mciteSetBstMidEndSepPunct{\mcitedefaultmidpunct}
{\mcitedefaultendpunct}{\mcitedefaultseppunct}\relax
\EndOfBibitem
\bibitem[Kullback and Leibler(1951)Kullback, and Leibler]{kl:ams51}
Kullback,~S.; Leibler,~R.~A. \emph{Ann. Math. Stat.} \textbf{1951}, \emph{22},
  79--86\relax
\mciteBstWouldAddEndPuncttrue
\mciteSetBstMidEndSepPunct{\mcitedefaultmidpunct}
{\mcitedefaultendpunct}{\mcitedefaultseppunct}\relax
\EndOfBibitem
\bibitem[Foster et~al.(1997)Foster, Wuttke, Radhakrishnan, Case, Gottesfeld,
  and Wright]{wright:nsb97}
Foster,~M.~P.; Wuttke,~D.~S.; Radhakrishnan,~I.; Case,~D.~A.;
  Gottesfeld,~J.~M.; Wright,~P.~E. \emph{Nat. Struct. Biol.} \textbf{1997},
  \emph{4}, 605--608\relax
\mciteBstWouldAddEndPuncttrue
\mciteSetBstMidEndSepPunct{\mcitedefaultmidpunct}
{\mcitedefaultendpunct}{\mcitedefaultseppunct}\relax
\EndOfBibitem
\bibitem[{MacKerell,~Jr.} et~al.(1998){MacKerell,~Jr.}, Bashford, Bellott,
  {Dunbrack,~Jr.}, Evanseck, Field, Fischer, Gao, Guo, Ha, Joseph-McCarthy,
  Kuchnir, Kuczera, Lau, Mattos, Michnick, Ngo, Nguyen, Prodhom, {Reiher,~III},
  Roux, Scklenkrich, Smith, Stote, Straub, Watanabe, Wi{\'o}rkiewicz-Kuczera,
  Yin, and Karplus]{charmm:jpcb98}
{MacKerell,~Jr.},~A.~D. et~al.  \emph{J. Phys. Chem. B} \textbf{1998},
  \emph{102}, 3586--3616\relax
\mciteBstWouldAddEndPuncttrue
\mciteSetBstMidEndSepPunct{\mcitedefaultmidpunct}
{\mcitedefaultendpunct}{\mcitedefaultseppunct}\relax
\EndOfBibitem
\bibitem[Foloppe et~al.(2001)Foloppe, Sagemark, Nordstrand, Berndt, and
  Nilsson]{nilsson:thiolate01}
Foloppe,~N.; Sagemark,~J.; Nordstrand,~K.; Berndt,~K.~D.; Nilsson,~L. \emph{J.
  Mol. Biol.} \textbf{2001}, \emph{310}, 449--470\relax
\mciteBstWouldAddEndPuncttrue
\mciteSetBstMidEndSepPunct{\mcitedefaultmidpunct}
{\mcitedefaultendpunct}{\mcitedefaultseppunct}\relax
\EndOfBibitem
\bibitem[Calimet and T.Simonson(2006)Calimet, and T.Simonson]{simonson:jmgm06}
Calimet,~N.; T.Simonson, \emph{J. Mol. Graph. Model.} \textbf{2006}, \emph{24},
  404--411\relax
\mciteBstWouldAddEndPuncttrue
\mciteSetBstMidEndSepPunct{\mcitedefaultmidpunct}
{\mcitedefaultendpunct}{\mcitedefaultseppunct}\relax
\EndOfBibitem
\bibitem[Phillips et~al.(2005)Phillips, Braun, Wang, Gumbart, Tajkhorshid,
  Villa, Chipot, Skeel, Kale, and Schulten]{namd:cc05}
Phillips,~J.; Braun,~R.; Wang,~W.; Gumbart,~J.; Tajkhorshid,~E.; Villa,~E.;
  Chipot,~C.; Skeel,~R.; Kale,~L.; Schulten,~K. \emph{J. Comp. Chem.}
  \textbf{2005}, \emph{26}, 1781--1802\relax
\mciteBstWouldAddEndPuncttrue
\mciteSetBstMidEndSepPunct{\mcitedefaultmidpunct}
{\mcitedefaultendpunct}{\mcitedefaultseppunct}\relax
\EndOfBibitem
\bibitem[Pang(2001)]{pang:psfg01}
Pang,~Y.~P. \emph{Proteins: Struc. Func. Bioinfo.} \textbf{2001}, \emph{45},
  183--189\relax
\mciteBstWouldAddEndPuncttrue
\mciteSetBstMidEndSepPunct{\mcitedefaultmidpunct}
{\mcitedefaultendpunct}{\mcitedefaultseppunct}\relax
\EndOfBibitem
\bibitem[Jorgensen et~al.(1983)Jorgensen, Chandrasekhar, Madura, Impey, and
  Klein]{tip32}
Jorgensen,~W.; Chandrasekhar,~J.; Madura,~J.~D.; Impey,~R.~W.; Klein,~M.~L.
  \emph{J. Chem. Phys.} \textbf{1983}, \emph{79}, 926--935\relax
\mciteBstWouldAddEndPuncttrue
\mciteSetBstMidEndSepPunct{\mcitedefaultmidpunct}
{\mcitedefaultendpunct}{\mcitedefaultseppunct}\relax
\EndOfBibitem
\bibitem[Neria et~al.(1996)Neria, Fischer, and Karplus]{tip3mod}
Neria,~E.; Fischer,~S.; Karplus,~M. \emph{J. Chem. Phys.} \textbf{1996},
  \emph{105}, 1902--1921\relax
\mciteBstWouldAddEndPuncttrue
\mciteSetBstMidEndSepPunct{\mcitedefaultmidpunct}
{\mcitedefaultendpunct}{\mcitedefaultseppunct}\relax
\EndOfBibitem
\bibitem[Humphrey et~al.(1996)Humphrey, Dalke, and Schulten]{vmd:jmg96}
Humphrey,~W.; Dalke,~A.; Schulten,~K. \emph{J. Mol. Graphics} \textbf{1996},
  \emph{14}, 33--38\relax
\mciteBstWouldAddEndPuncttrue
\mciteSetBstMidEndSepPunct{\mcitedefaultmidpunct}
{\mcitedefaultendpunct}{\mcitedefaultseppunct}\relax
\EndOfBibitem
\bibitem[Feller et~al.(1995)Feller, Zhang, Pastor, and Brooks]{feller:jcp95}
Feller,~S.; Zhang,~Y.; Pastor,~R.; Brooks,~B. \emph{J. Chem. Phys.}
  \textbf{1995}, \emph{103}, 4613--4621\relax
\mciteBstWouldAddEndPuncttrue
\mciteSetBstMidEndSepPunct{\mcitedefaultmidpunct}
{\mcitedefaultendpunct}{\mcitedefaultseppunct}\relax
\EndOfBibitem
\bibitem[Stewart(1989)]{stuart:jcc89a}
Stewart,~J. J.~P. \emph{J. Comp. Chem.} \textbf{1989}, \emph{10},
  209--220\relax
\mciteBstWouldAddEndPuncttrue
\mciteSetBstMidEndSepPunct{\mcitedefaultmidpunct}
{\mcitedefaultendpunct}{\mcitedefaultseppunct}\relax
\EndOfBibitem
\bibitem[Stewart(1989)]{stuart:jcc89b}
Stewart,~J. J.~P. \emph{J. Comp. Chem.} \textbf{1989}, \emph{10},
  221--264\relax
\mciteBstWouldAddEndPuncttrue
\mciteSetBstMidEndSepPunct{\mcitedefaultmidpunct}
{\mcitedefaultendpunct}{\mcitedefaultseppunct}\relax
\EndOfBibitem
\bibitem[Frisch et~al.()Frisch, Trucks, Schlegel, Scuseria, Robb, Cheeseman,
  Scalmani, Barone, Mennucci, Petersson, Nakatsuji, Caricato, Li, Hratchian,
  Izmaylov, Bloino, Zheng, Sonnenberg, Hada, Ehara, Toyota, Fukuda, Hasegawa,
  Ishida, Nakajima, Honda, Kitao, Nakai, Vreven, Montgomery, Peralta, Ogliaro,
  Bearpark, Heyd, Brothers, Kudin, Staroverov, Kobayashi, Normand,
  Raghavachari, Rendell, Burant, Iyengar, Tomasi, Cossi, Rega, Millam, Klene,
  Knox, Cross, Bakken, Adamo, Jaramillo, Gomperts, Stratmann, Yazyev, Austin,
  Cammi, Pomelli, Ochterski, Martin, Morokuma, Zakrzewski, Voth, Salvador,
  Dannenberg, Dapprich, Daniels, Farkas, Foresman, Ortiz, Cioslowski, and
  Fox]{g09}
Frisch,~M.~J. et~al.  Gaussian~09 {R}evision {A}.1. Gaussian Inc. Wallingford
  CT 2009\relax
\mciteBstWouldAddEndPuncttrue
\mciteSetBstMidEndSepPunct{\mcitedefaultmidpunct}
{\mcitedefaultendpunct}{\mcitedefaultseppunct}\relax
\EndOfBibitem
\bibitem[Schmidt et~al.(1993)Schmidt, Baldridge, Boatz, Elbert, Gordon, Jensen,
  Koseki, Matsunaga, Nguyen, Su, Windus, Dupuis, and Montgomery]{gamess}
Schmidt,~M.~W.; Baldridge,~K.~K.; Boatz,~J.~A.; Elbert,~S.~T.; Gordon,~M.~S.;
  Jensen,~J.~J.; Koseki,~S.; Matsunaga,~N.; Nguyen,~K.~A.; Su,~S.;
  Windus,~T.~L.; Dupuis,~M.; Montgomery,~J.~A. \emph{J. Comp. Chem.}
  \textbf{1993}, \emph{14}, 1347--1363\relax
\mciteBstWouldAddEndPuncttrue
\mciteSetBstMidEndSepPunct{\mcitedefaultmidpunct}
{\mcitedefaultendpunct}{\mcitedefaultseppunct}\relax
\EndOfBibitem
\bibitem[Berg and Merkle(1989)Berg, and Merkle]{berg:jacs89}
Berg,~J.~M.; Merkle,~D.~L. \emph{J. Am. Chem. Soc.} \textbf{1989}, \emph{111},
  3759--3761\relax
\mciteBstWouldAddEndPuncttrue
\mciteSetBstMidEndSepPunct{\mcitedefaultmidpunct}
{\mcitedefaultendpunct}{\mcitedefaultseppunct}\relax
\EndOfBibitem
\bibitem[Krizek and Berg(1992)Krizek, and Berg]{berg:ic92}
Krizek,~B.~A.; Berg,~J.~M. \emph{Inorg. Chem.} \textbf{1992}, \emph{31},
  2984--2986\relax
\mciteBstWouldAddEndPuncttrue
\mciteSetBstMidEndSepPunct{\mcitedefaultmidpunct}
{\mcitedefaultendpunct}{\mcitedefaultseppunct}\relax
\EndOfBibitem
\bibitem[Asthagiri et~al.(2004)Asthagiri, Pratt, Paulaitis, and
  Rempe]{asthagiri:jacs04}
Asthagiri,~D.; Pratt,~L.~R.; Paulaitis,~M.~E.; Rempe,~S.~B. \emph{J. Am. Chem.
  Soc.} \textbf{2004}, \emph{126}, 1285--1289\relax
\mciteBstWouldAddEndPuncttrue
\mciteSetBstMidEndSepPunct{\mcitedefaultmidpunct}
{\mcitedefaultendpunct}{\mcitedefaultseppunct}\relax
\EndOfBibitem
\bibitem[Shi et~al.(1992)Shi, Beger, and Berg]{berg:bj93}
Shi,~Y.; Beger,~R.~D.; Berg,~J.~M. \emph{Biophys. J.} \textbf{1992}, \emph{64},
  749--753\relax
\mciteBstWouldAddEndPuncttrue
\mciteSetBstMidEndSepPunct{\mcitedefaultmidpunct}
{\mcitedefaultendpunct}{\mcitedefaultseppunct}\relax
\EndOfBibitem
\bibitem[Miura et~al.(1998)Miura, Satoh, and Takeuchi]{takeuchi:bba98}
Miura,~T.; Satoh,~T.; Takeuchi,~H. \emph{Biochim. Biophys. Acta} \textbf{1998},
  \emph{1384}, 171--179\relax
\mciteBstWouldAddEndPuncttrue
\mciteSetBstMidEndSepPunct{\mcitedefaultmidpunct}
{\mcitedefaultendpunct}{\mcitedefaultseppunct}\relax
\EndOfBibitem
\bibitem[Hanas et~al.(2005)Hanas, Larabee, and Hocker]{hanas:2005}
Hanas,~J.~S.; Larabee,~J.~L.; Hocker,~J.~R. In \emph{{Zinc Finger Proteins:
  From} atomic contact to cellular Function}; Iuchi,~S., Kuldell,~N., Eds.;
  Springer series on Life and Biomedical Sciences; Kluwer Academic/Plenum
  Publishers: Boston, 2005; Chapter 8\relax
\mciteBstWouldAddEndPuncttrue
\mciteSetBstMidEndSepPunct{\mcitedefaultmidpunct}
{\mcitedefaultendpunct}{\mcitedefaultseppunct}\relax
\EndOfBibitem
\bibitem[Nomura and Sugiura(2002)Nomura, and Sugiura]{sugiura:ic02}
Nomura,~A.; Sugiura,~Y. \emph{Inorg. Chem.} \textbf{2002}, \emph{41},
  3693--3698\relax
\mciteBstWouldAddEndPuncttrue
\mciteSetBstMidEndSepPunct{\mcitedefaultmidpunct}
{\mcitedefaultendpunct}{\mcitedefaultseppunct}\relax
\EndOfBibitem
\bibitem[Krizek et~al.(1991)Krizek, Amann, Kilfoil, Merkle, and
  Berg]{Krizek:jacs91}
Krizek,~B.~A.; Amann,~B.~T.; Kilfoil,~V.~J.; Merkle,~D.~L.; Berg,~J.~M.
  \emph{J. Am. Chem. Soc.} \textbf{1991}, \emph{113}, 4518--4523\relax
\mciteBstWouldAddEndPuncttrue
\mciteSetBstMidEndSepPunct{\mcitedefaultmidpunct}
{\mcitedefaultendpunct}{\mcitedefaultseppunct}\relax
\EndOfBibitem
\bibitem[Merchant and Asthagiri(2009)Merchant, and Asthagiri]{merchant:jcp09}
Merchant,~S.; Asthagiri,~D. \emph{J. Chem. Phys} \textbf{2009}, \emph{130},
  195102\relax
\mciteBstWouldAddEndPuncttrue
\mciteSetBstMidEndSepPunct{\mcitedefaultmidpunct}
{\mcitedefaultendpunct}{\mcitedefaultseppunct}\relax
\EndOfBibitem
\bibitem[Asthagiri et~al.(2003)Asthagiri, Pratt, and Ashbaugh]{lrp:ionsjcp03}
Asthagiri,~D.; Pratt,~L.~R.; Ashbaugh,~H.~S. \emph{J. Chem. Phys.}
  \textbf{2003}, \emph{119}, 2702--2708\relax
\mciteBstWouldAddEndPuncttrue
\mciteSetBstMidEndSepPunct{\mcitedefaultmidpunct}
{\mcitedefaultendpunct}{\mcitedefaultseppunct}\relax
\EndOfBibitem
\bibitem[Fatmi et~al.(2005)Fatmi, Hofer, Randolf, and Rode]{rode:znjcp}
Fatmi,~M.~Q.; Hofer,~T.~S.; Randolf,~B.~R.; Rode,~B.~M. \emph{J. Chem. Phys.}
  \textbf{2005}, \emph{123}, 054514\relax
\mciteBstWouldAddEndPuncttrue
\mciteSetBstMidEndSepPunct{\mcitedefaultmidpunct}
{\mcitedefaultendpunct}{\mcitedefaultseppunct}\relax
\EndOfBibitem
\bibitem[Lachenmann et~al.(2004)Lachenmann, Ladbury, Dong, Huang, Carey, and
  Weiss]{weiss:biochem04}
Lachenmann,~M.~J.; Ladbury,~J.~E.; Dong,~J.; Huang,~K.; Carey,~P.; Weiss,~M.~A.
  \emph{Biochemistry} \textbf{2004}, \emph{43}, 13910--13925\relax
\mciteBstWouldAddEndPuncttrue
\mciteSetBstMidEndSepPunct{\mcitedefaultmidpunct}
{\mcitedefaultendpunct}{\mcitedefaultseppunct}\relax
\EndOfBibitem
\bibitem[Sakharov and Lim(2009)Sakharov, and Lim]{lim:jacs09}
Sakharov,~D.~V.; Lim,~C. \emph{J. Am. Chem. Soc.} \textbf{2009}, \emph{30-2},
  191--202\relax
\mciteBstWouldAddEndPuncttrue
\mciteSetBstMidEndSepPunct{\mcitedefaultmidpunct}
{\mcitedefaultendpunct}{\mcitedefaultseppunct}\relax
\EndOfBibitem
\bibitem[Hummer et~al.(1998)Hummer, Pratt, and Garcia]{Hummer:jpca98}
Hummer,~G.; Pratt,~L.~R.; Garcia,~A.~E. \emph{J. Phys. Chem. A.} \textbf{1998},
  \emph{102}, 7885--7895\relax
\mciteBstWouldAddEndPuncttrue
\mciteSetBstMidEndSepPunct{\mcitedefaultmidpunct}
{\mcitedefaultendpunct}{\mcitedefaultseppunct}\relax
\EndOfBibitem
\end{mcitethebibliography}

\providecommand*\mcitethebibliography{\thebibliography}
\csname @ifundefined\endcsname{endmcitethebibliography}
  {\let\endmcitethebibliography\endthebibliography}{}

\end{document}